\documentclass[11pt]{emulateapj}
\usepackage{graphicx,amssymb,amsmath,times}
\usepackage{natbib,graphicx}
\usepackage[letterpaper]{hyperref}
\slugcomment{The Astrophysical Journal}
\newcommand{\flux}{\ensuremath{\mathrm{~erg~cm^{-2}~s^{-1}} }}

\def\flux{erg cm$^{-2}$ s$^{-1}$}
\def\latflux{ph cm$^{-2}$ s$^{-1}~$}

\newcommand{\fermi}{\emph{Fermi}~}

\def\rxte{\emph{RXTE}~}
\def\swift{\emph{Swift}~}
\begin{document}
%
\title{\vspace{-0.3cm}The first Fermi multifrequency campaign on BL Lacertae:\\ characterizing the low-activity state of the eponymous blazar}
%
\slugcomment{Accepted by The Astrophysical Journal}
%
\shorttitle{The first Fermi multifrequency campaign on BL Lac}
\shortauthors{Abdo et~al.}
%
%
%
%

\author{ %
\footnotesize{
A.~A.~Abdo\altaffilmark{1,2},
M.~Ackermann\altaffilmark{3},
M.~Ajello\altaffilmark{3},
E.~Antolini\altaffilmark{4,5*},
L.~Baldini\altaffilmark{7},
J.~Ballet\altaffilmark{8},
G.~Barbiellini\altaffilmark{9,10},
D.~Bastieri\altaffilmark{11,12},
K.~Bechtol\altaffilmark{3},
R.~Bellazzini\altaffilmark{7}
% ...
} 
}
%
%
%
\altaffiltext{*}{Corresponding authors: S.~Ciprini, stefano.ciprini@pg.infn.it; E.~Antolini, elisa.antolini@tiscali.it;  D.~Donato, davide.donato-1@nasa.gov; K.~V.~Sokolovsky, ksokolov@mpifr.de; G.~Tosti, gino.tosti@pg.infn.it.}
\altaffiltext{1}{Space Science Division, Naval Research Laboratory, Washington, DC 20375, USA}
\altaffiltext{2}{National Research Council Research Associate, National Academy of Sciences, Washington, DC 20001, USA}
\altaffiltext{3}{W. W. Hansen Experimental Physics Laboratory, Kavli Institute for Particle Astrophysics and Cosmology, Department of Physics and SLAC National Accelerator Laboratory, Stanford University, Stanford, CA 94305, USA}
\altaffiltext{4}{Istituto Nazionale di Fisica Nucleare, Sezione di Perugia, I-06123 Perugia, Italy}
\altaffiltext{5}{Dipartimento di Fisica, Universit\`a degli Studi di Perugia, I-06123 Perugia, Italy}
\altaffiltext{7}{Istituto Nazionale di Fisica Nucleare, Sezione di Pisa, I-56127 Pisa, Italy}
\altaffiltext{8}{Laboratoire AIM, CEA-IRFU/CNRS/Universit\'e Paris Diderot, Service d'Astrophysique, CEA Saclay, 91191 Gif sur Yvette, France}
\altaffiltext{9}{Istituto Nazionale di Fisica Nucleare, Sezione di Trieste, I-34127 Trieste, Italy}
\altaffiltext{10}{Dipartimento di Fisica, Universit\`a di Trieste, I-34127 Trieste, Italy}
\altaffiltext{11}{Istituto Nazionale di Fisica Nucleare, Sezione di Padova, I-35131 Padova, Italy}
\altaffiltext{12}{Dipartimento di Fisica ``G. Galilei", Universit\`a di Padova, I-35131 Padova, Italy}
% ...
%
%
\begin{abstract}
%
%
\footnotesize{
We report on observations of BL~Lacertae during the first 18 months of \fermi\-LAT science operations and present results from a 48-day multifrequency coordinated campaign from 2008 August 19 to 2008 October 7. The radio to gamma-ray behavior of BL~Lac is unveiled during a low activity state thanks to the coordinated observations of radio-band (Mets\"ahovi and VLBA), near-IR/optical (Tuorla, Steward, OAGH and MDM) and X-ray (\rxte\ and \swift) observatories. No variability was resolved in gamma-rays during the campaign, and the brightness level was 15 times lower than the level of the 1997 EGRET outburst.  Moderate and uncorrelated variability has been detected in UV and X-rays. The X-ray spectrum is found to be concave indicating the transition region between the low and high energy component of the spectral energy distribution (SED). VLBA observation detected a synchrotron  spectrum
self-absorption turnover in the innermost part of the radio jet appearing to be elongated and inhomogeneous, and constrained the average magnetic field there to be less than 3 G. Over the following months BL~Lac appeared variable in gamma-rays, showing flares (in 2009 April and 2010 January). There is no evidence for correlation of the gamma-rays with the optical flux monitored from the ground in 18 months. The SED may be described by a single zone or two zone synchrotron self-Compton (SSC) model, but a hybrid SSC plus external radiation Compton (ERC) model seems preferred based on the  observed variability and the fact that it provides a fit closest to equipartition.
}
\end{abstract}
\keywords{Gamma rays: galaxies -- BL Lacertae objects: individual: \object{BL~Lac} -- BL Lacertae objects: general
-- X-rays: galaxies -- galaxies: jets -- radiation mechanisms: non-thermal}
%
%
%
%
\section{Introduction}\label{sect:introduction}  %
%
%
BL~Lacertae (BL~Lac, S4 2200+42, OY 401, B3 2200+420, 1ES~2200+420,
3EG~J2202+4217, 1FGL~J2202.8+4216) at redshift $z=0.0686$
\citep[e.g.,][]{vermeulen95} was historically the prototype of the class of active galactic nuclei (AGN) for which BL Lac has become the eponym. It is categorized as an intermediate synchrotron peaked
(ISP) BL~Lac object.  This is based on the latest HSP/ISP/LSP
classification proposed by \citet{LAT_seds} to replace the older
HBL-IBL-LBL scheme \citep[][]{padovani95,fossati98}. BL~Lac
occasionally showed peculiar behavior that has questioned both this classification and a simple interpretation of its broad-band emission
in terms of synchrotron and synchrotron self-Compton (SSC) emission
produced by a single blob. In fact, BL~Lac is a source that has shown
quite complex and distinct X-ray spectral behavior
\citep{madejski99,ravasio02,boettcher04} and during several epochs
broad H$\alpha$ and H$\beta$ emission lines with luminosity ($\sim
10^{41} \rm \, erg \, s^{-1}$) comparable to those of type I Seyfert
galaxies \citep{vermeulen95,corbett96,corbett00}. There is
evidence for an increase by $\sim50\%$ in ten years of the broad
H$\alpha$ line and an underluminous broad line region (BLR), compared
to other AGN.  The narrow lines and radio luminosities of BL~Lac match
those of of low-excitation and miniature radio galaxies
\citep{capetti10}. Long term radio-optical flux density monitoring
and several multiwavelength campaigns have been carried out on BL~Lac, like the past campaigns of the Whole Earth Blazar Telescope \citep[WEBT,][]{villata02a,villata02b} dedicated to this source,
providing very complete datasets \citep[for
example,][]{bloom97,sambruna99,madejski99,ravasio03,boettcher03,villata02a,villata04a,villata04b,villata09,raiteri09,raiteri10,marscher08}.
In particular in summer 1997 BL Lac showed the largest optical outburst ever recorded in almost 33 years \citep{nesci98,tosti99,villata04b}.

In the X-ray band the two broad components of the spectral energy
distribution (SED) are overlapping and the radiation at this band is
at times dominated by the high-energy end of the synchrotron emission,
while at other occasions it is dominated by the low-frequency portion
of the high-energy component
\citep[e.g.,][]{madejski99,ravasio02}. The rapid X-ray variability is
mainly restricted to the low-energy excess portion of the X-ray
spectrum, presumably produced by synchrotron radiation
\citep{ravasio02,ravasio03}.  The X-ray variability of BL~Lac,
except for a few major flaring events, has a log-normal distribution
\citep{giebels09}, meaning that the emission is a multiplicative
product of a large number of independent random events.

\begin{figure}[t!!]
\centering \hskip -1.4cm%
\resizebox{10cm}{!}{\rotatebox[]{0}{\includegraphics{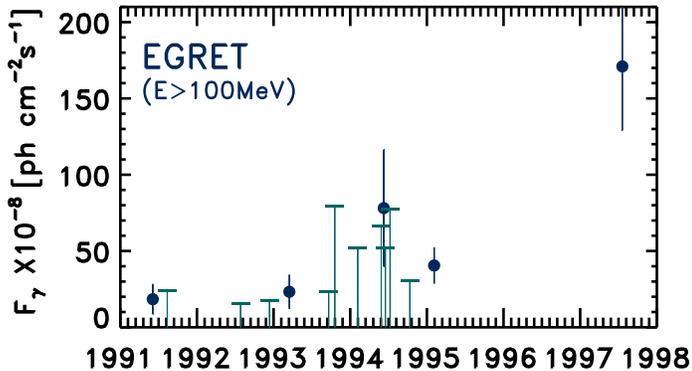}}}
\vskip -0.5cm
\caption{Gamma-ray light curve of BL~Lac obtained by EGRET in the period 1991-1997. Data points have source detection significance above 4$\sigma$ \citep{nandikotkur07}.
} \label{fig:EGRET_lc}
\end{figure}

Gamma-ray observations by EGRET resulted in several detections (Figure
\ref{fig:EGRET_lc}). Observations after 1995 resulted in an average
$\gamma$-ray flux above 100 MeV of $(40 \pm 12) \times 10^{-8}$
\latflux \citep{catanese97} and in a flare in 1997 at a level $(171
\pm 42) \times 10^{-8}$ \latflux with 10.2$\sigma$ significance
\citep{bloom97}. Correlated $\gamma$-rays and optical flaring emission
were observed during the EGRET era \citep{bloom97}.  Very high energy
(VHE) $\gamma$-ray emission was claimed by the Crimean GT-48
\citep{neshpor01} and HEGRA \citep{kranich03} atmospheric Cherenkov
telescopes. A significant ($>5\ \sigma$) detection above 200 GeV was made only by the MAGIC telescope in 2005 \citep{albert07}. Gamma-ray emission was explained
with the requirement of Comptonization of external-jet photons
(external-radiation Compton, ERC) in addition to the SSC (in-jet)
emission \citep{madejski99,boettcher00}. Superluminal motion
of $\beta_{\rm app}$ up to $(10.57 \pm 0.74)$ has been observed in
this object \citep{lister09,denn00}. Following results of past
multifrequency campaigns, possibly distinct VLBI jet structures are
assumed to contribute, sometimes with delays, to the radio flux
density light curves, and sometimes are suggested to be responsible
for optical to TeV $\gamma$-ray flares \citep[for
example,][]{bach06,marscher08}.

%
%

In this paper, we report on LAT observations of BL~Lac during the
first 18 months of \fermi\ science operations (from 2008 August 04 to
2010 February 04).  The LAT's reasonably uniform exposure, high
sensitivity, and continuous sky monitoring make it an excellent
instrument around which to organize a simultaneous multifrequency
campaign. The so-called \fermi\ planned intensive campaign (PIC)
dedicated to BL~Lac was performed in 2008 August 19 -- 2008 October 7
(MJD 54697.8 -- 54746), during roughly the first two months of science
operations, irrespective of the brightness of the source.  This was
part of a series of \emph{Fermi}-LAT-collaboration PICs
\citep{tosti07,thompson07}, involving observing proposals submitted to
the \rxte\ (Cycle 12) and \swift\ (Cycle 4) X-ray satellites, and
organized in advance of the \fermi\ launch. This ensured access to the
facilities allowing the best multiwavelength (MW) coverage.  The aim
of the campaign is to shed light for the first time on the broad-band
radio-to-$\gamma$-ray SED, including the high energy (X-ray and
$\gamma$-ray) behavior, during a low activity phase of the source. The
18-month LAT light curve shows that BL~Lac was variable in
$\gamma$-rays for most of these 18 months, with the exception of the
first 2 months, which corresponds to the period of the MW campaign. In
section \ref{sect:LAT} the light curve by \fermi-LAT during the first
18 months of survey and simultaneous optical long-term monitoring
data are presented joined to the analysis of the $\gamma$-ray
spectrum. In section \ref{sect:MultifrequencyCampaign} the
multiwaveband radio-band to X-ray flux density and parsec-scale
radio structure observations collected during the campaign period are reported
and discussed. The assembled spectral energy distribution and modeling
are reported in section \ref{sect:mwsed} and conclusions in section
\ref{sect:conclusion}.

A $\Lambda$CDM cosmology with values given
within 1$\sigma$ of the WMAP results \citep{komatsu09}, namely
$\Omega_m = 0.27$, $\Omega_\Lambda = 0.73$, and $H_0=71$ km s$^{-1}$
Mpc$^{-1}$\ is used.

%
%
%
%
\section{Gamma-ray observations and results by Fermi-LAT}\label{sect:LAT}
%
\subsection{LAT analysis and observations}\label{subsect:LATobs}
%
%
%
%
\begin{figure*}[tt!!]
\centering \hskip -0.3cm%
\resizebox{\hsize}{!}{\rotatebox[]{0}{\includegraphics{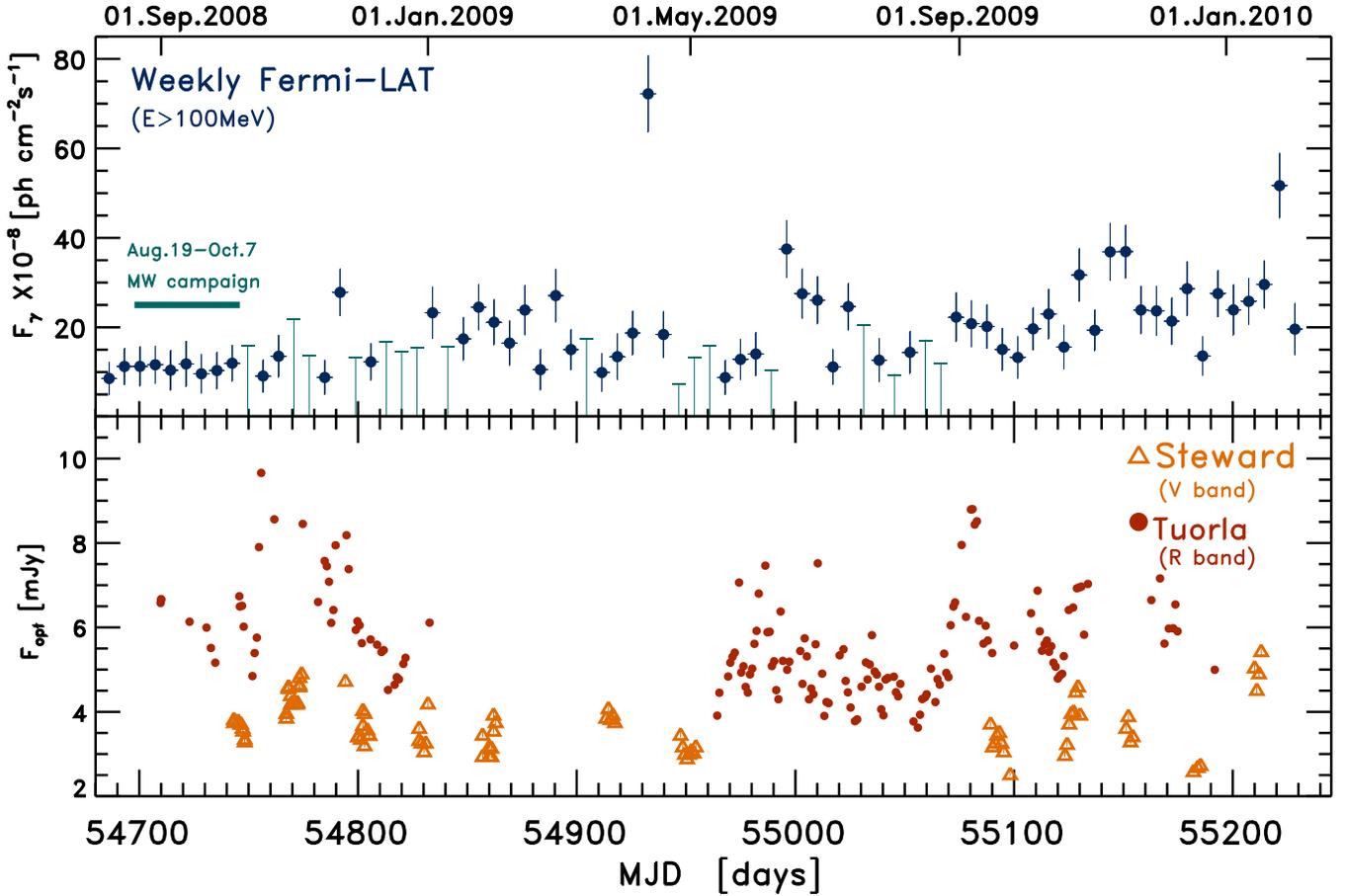}}}
\vskip -0.2cm \caption{Top panel: 18-month weekly integrated flux ($E>100$MeV) light curve of BL~Lac obtained obtained from 2008, Aug. 04 to 2010, Feb. 04. Bottom panel: superposed optical V-band and R-band observed flux (not corrected for interstellar absorption and host galaxy contribution) light curves obtained respectively at the Steward Observatory, Kitt Peak, Arizona, USA, and Tuorla Observatory, Turku, Finland, in the same period.
}
\label{fig:18month_LAT_tuorla}
\end{figure*}
%
%

The Large Area Telescope (LAT), on board the \fermi\ Gamma-ray
Space Telescope \citep{atwood09}, is a pair-conversion $\gamma$-ray
telescope, sensitive to photon energies from about 20 MeV up to $>$300
GeV. It consists of a tracker (composed of two sections, front and back, with different angular resolutions), a calorimeter and an
anti-coincidence system to reject the charged-particle
background. The LAT has a large peak effective area ($\sim
8000$~cm$^2$ for 1 GeV photons in the event class considered here),
viewing $\approx 2.4$~sr of the sky with angular resolution (68\%
containment radius) better than $ \approx 1^\circ$ at $E = 1$~GeV. The
large field of view, improved effective area and sensitivity and the
survey nominal mode make \emph{Fermi}-LAT an optimal all-sky hunter for
high-energy flares and an unprecedented monitor of $\gamma$-ray
sources.
The data set used in this paper was collected during the first 18
months of \fermi\ science obserations, from 2008 August 4 to 2010
February 4 (about 550 days, 78 weeks from MJD 54682.7 to 55232.9, as
shown in the weekly light curve in Figure
\ref{fig:18month_LAT_tuorla}, upper panel). This interval includes the
period chosen for the first \fermi\ MW campaign on BL~Lac (2008, Aug.\
19 -- Oct.\ 7, MJD 54697.8-54746, i.e. about 48 days, indicated by the
line in Figure \ref{fig:18month_LAT_tuorla}, upper panel). \fermi-LAT
data analysis was performed with the standard \fermi-LAT
\texttt{ScienceTools} software
package\footnote{\url{http://fermi.gsfc.nasa.gov/ssc/data/analysis/documentation/Cicerone/}}
using version v9r15p5. Only events belonging to the ``Diffuse'' class
in the energy range $0.1-100$ GeV were used in the
analysis. Instrument response functions (IRFs) used were
P6\_V3\_DIFFUSE. In order to provide protection against significant
background contamination by Earth-limb $\gamma$ rays, all events with
zenith angles $> 105^\circ$ were excluded.

The 18-month light curve was built using 1-week time bins (Figures
\ref{fig:18month_LAT_tuorla} and
\ref{fig:multipanelMWlightcurves}). For each time bin the integrated
flux (E$>$100 MeV) values were computed using the maximum-likelihood
algorithm implemented in the science tool \texttt{gtlike}. For each
time bin we analyzed a region of interest (RoI) of $12^{\circ}$ in
radius, centered on the position of the source.
All point sources listed in the 1FGL (1-year) LAT catalog
\citep{1FGLcatalog} within $19^{\circ}$ from BL~Lacertae and having
test statistic\footnote{The test statistic is defined as
$TS=2\Delta$log(Likelihood) between models with and without the source
and it is a measure of the source significance
\citep{mattox96}.}  TS $>50$ and fluxes above $10^{-8}$
\latflux were included in the RoI model using a power-law spectrum
($dN/dE \propto E^{-\Gamma}$, where $\Gamma$ is the photon index).
The isotropic background (the sum of the residual instrumental
background and extragalactic diffuse $\gamma$-ray background) was
included in the RoI model using the standard model file
\texttt{isotropic\_iem\_v02.txt}\footnote{\url{http://fermi.gsfc.nasa.gov/ssc/data/access/lat/BackgroundModels.html}},
and the Galactic diffuse emission was included in the modeling using
the file and the standard file \texttt{gll\_iem\_v02.fit}.  In the
final light curve computation the photon index value was frozen to the
value resulting from the likelihood analysis on the entire
period. For each time bin, if the TS value for the source was $TS<4$
or the number of model predicted photons $N_{pred} <10$, the value of
the fluxes were replaced by the 2-$\sigma$ upper limits. All errors
reported in the figures or quoted in the text are 1-$\sigma$
statistical errors. The estimated relative systematic uncertainty on the
flux, according to \citet{1FGLcatalog} and reflecting the relative
systematic uncertainty on effective area, is set to 10\% at 100 MeV, 5\% at 500 MeV and 20\% at 10 GeV.
%
%
\begin{figure}[t!!]
\centering \hskip -0.8cm%
\resizebox{8.4cm}{!}{\rotatebox[]{-90}{\includegraphics{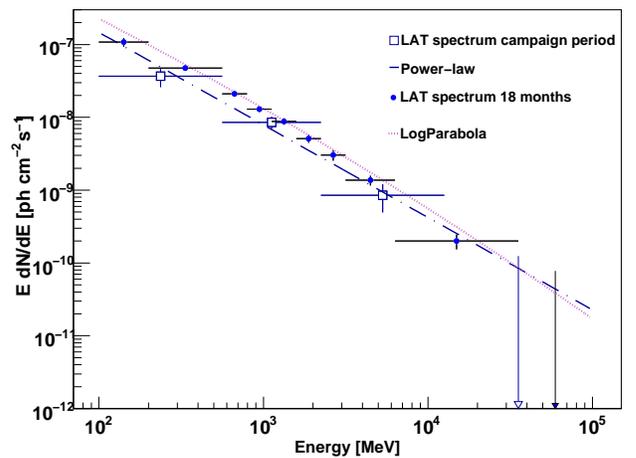}}}
\vskip -1.0cm
\caption{18-month averaged energy spectrum of BL~Lac obtained with a band likelihood over the accumulated observations from 2008 Aug.\ 04 to 2010 Feb.\ 04 (MJD 54682.7-55232.9). The spectrum shows a departure from a simple power law that can be fitted by a log parabola function. The blue (open square) symbols represent the spectrum corresponding only to the 48-day period of the multifrequency campaign.
}
\label{fig:LAT18monthspec}
\end{figure}
%
%
\subsection{Gamma-ray spectral and temporal behavior}\label{subsect:LATresults}

The $\gamma$-ray spectral analysis of BL~Lacertae was performed both
for the entire 18-month period, and for the 48 days of the
campaign. These spectra are shown in Figure \ref{fig:LAT18monthspec}.
The flux value for each energy bin and for the two periods are
reported in Table \ref{table:spectrumpoints}. The most energetic
photon observed within the 95\% point spread function containment
radius of the source for the 18-month period had energy 70 GeV, while for the campaign period the highest energy photon had 10 GeV. The energy-spectrum binning was built requiring $TS > 50$
and/or model-predicted source photons $>8$, except for the last bin,
which had an upper limit rather than a detection.  The 18-month
spectrum results in an integrated average flux of ($19.85 \pm 0.40)
\times 10^{-8}$ \latflux, in the $0.1-100$ GeV range, with $TS=2375$
and photon index $\Gamma= 2.38 \pm 0.01$, while the value for the
48-day interval of the multifrequency campaign, ($11.5 \pm 2.7)
\times 10^{-8}$ \latflux, and the photon index slightly flatter
($\Gamma=2.27 \pm 0.10$), with a $TS = 111$.

In order to quantify the departure of the 18-month spectrum from
a power law shape, we use a spectral curvature index $\mathcal{C}$ as
in \citet{1FGLcatalog},
\begin{equation}
\mathcal{C} = \sum_{i} \frac{(F_{i}
- F_{i}^{PL})^{2}}{\sigma^{2} + (f_{i}^{rel} F_{i})^{2}}\ ,
\end{equation}
 where $i$ is the index for a particular energy bin, $F_{i}$ is the
observed flux, $F_{i}^{PL}$ is the flux in the $ith$ bin predicted by
the global power-law fit, $\sigma$ is the error in $F_{i}$, and
$f_{i}^{rel}$ is the relative systematic uncertainty in that bin
(reported in Table \ref{table:spectrumpoints}).  The power-law fit has
two parameters: the normalization and photon index.  The curvature
parameter is expected to follow a $\chi^{2}$ distribution with $9-2=7$
degrees of freedom (d.o.f.) if the power law hypothesis is true. With
99\% confidence the spectral shape is significantly different from a
power law when $\mathcal{C} > 18.48$. For the accumulated
18-month data on BL Lac we found $\mathcal{C} = 20.88$ ($0.1<E<100$
GeV), which implies a curved spectrum.  The same procedure, when
applied to the spectrum of the 48-day multifrequency campaign, resulted
in $\mathcal{C} = 2.83$ ($0.1<E<100$ GeV), implying there is no
significant evidence for deviation from a simple power-law in this
spectrum (Figures \ref{fig:LAT18monthspec} and \ref{fig:SED}).  This
does not necessarily mean that there is no curvature during this
48-day period, because it could be due to the lower emission state and
accumulated statistics.

To further explore curvature, the 18-month spectrum was fit with a
log-parabola function in the $0.1-300$ GeV energy band.  This was
performed in two runs: first a fit with $E_{break}$ left free, then another
with $E_{break}$ frozen at the value found in the first run (300
MeV) to improve the calculation of the other parameter
values.  The result of the latter fit gives
\begin{eqnarray}
\nonumber
dN/dE  &= & (2.02 \pm 0.07) \times 10^{-10}  \\ \nonumber
 & \times & (E/ 300\ MeV)^{- \left( (2.23 \pm 0.05) + (0.07\pm 0.02) \log(E/300\ MeV) \right)}\ ,
\end{eqnarray}
with TS = $ 2384.1$.
\begin{table}[t!!]
\scriptsize
\tabletypesize{\small}
\hskip -2cm
\caption{BL~Lac $\gamma$-ray spectrum obtained with likelihood analysis in each energy band.}\label{table:spectrumpoints}
%
\begin{tabular}{lll}
\hline \hline %
$E_{min}-E_{max}$ & $\gamma$-ray flux & Syst. \\
$[\mathrm{MeV}]$ & $[\times 10^{-8} \textrm{ph}~\textrm{cm}^{-2} \textrm{s}^{-1}]$ &  \\
\hline  %
 & 18 months  &  \\
 \hline  %
100-200 & 7.62$\pm$1.03 & 10\% \\
200-562 & 5.14$\pm$0.27 & 5\% \\
562-794 & 0.73$\pm$0.06 & 5\% \\
794-1122 & 0.45$\pm$0.04 & 5\% \\
1122-1585 & 0.31$\pm$0.03 & 10\% \\
1585-2239 & 0.18$\pm$0.02 & 10\%\\
2239-3162 & 0.11$\pm$0.02 & 10\% \\
3162-6310 & 0.10$\pm$0.01 & 15\% \\
6310-35481 & 0.04$\pm$0.01 & 15\% \\
\hline
 & Aug 19 - Sep 9 &  \\
\hline %
100-562 & 7.18$\pm$2.12 & 5\% \\
562-2239 & 1.28$\pm$0.24 & 10\% \\
2239-12589 & 0.17$\pm$0.07 & 15\% \\
\hline \hline %
\normalsize
\end{tabular}
\end{table}
%
The higher TS value shows that the 18 months LAT spectrum does not follow a
simple power law function, and the log-parabola model can
describe this spectrum.

\begin{figure*}[t!!]
\centering
\resizebox{16.5cm}{!}{\rotatebox[]{0}{\includegraphics{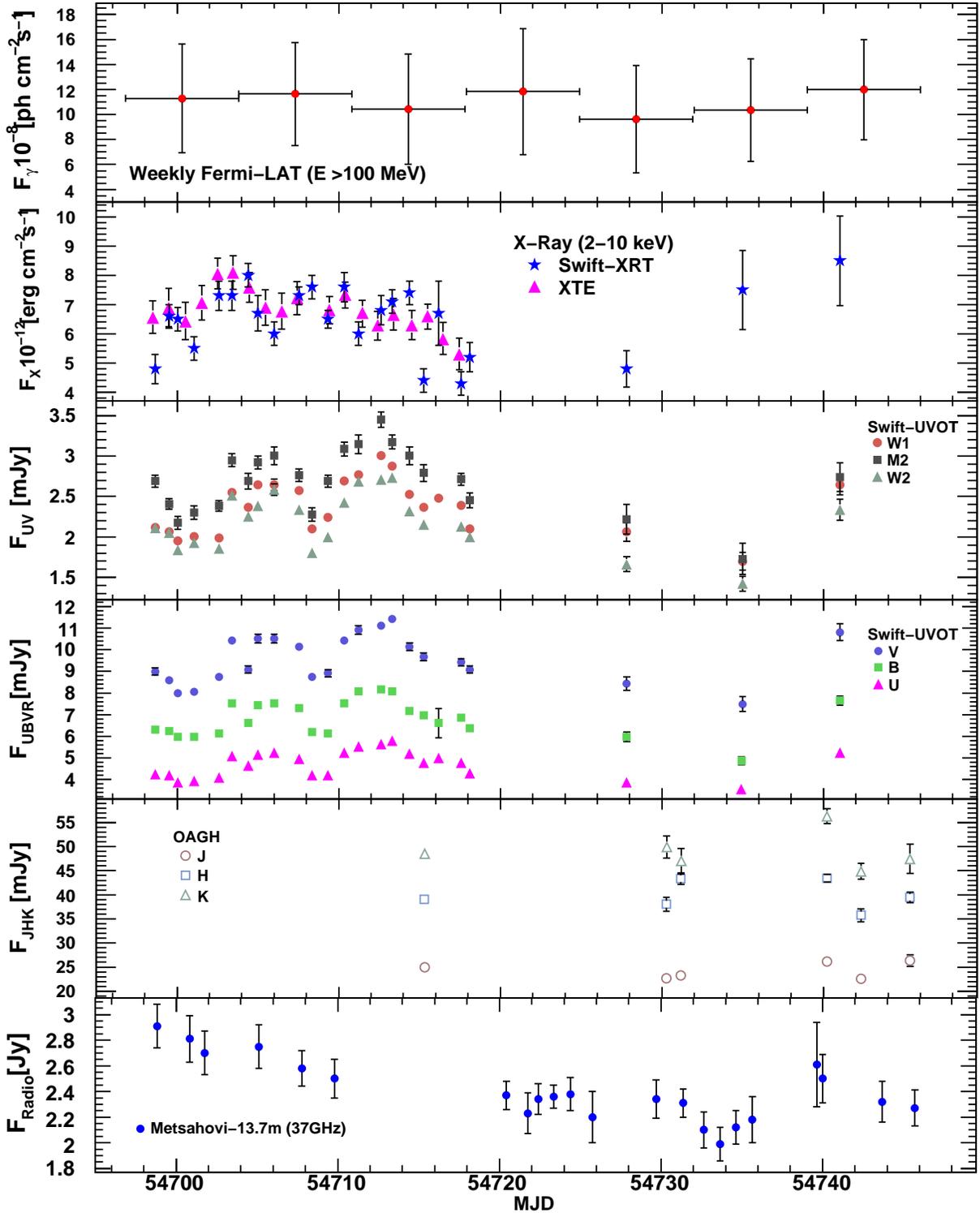}}}
\vskip -0.3cm \caption{Simultaneous multifrequency flux light curves of BL~Lac from the 48-days \fermi intensive campaign obtained with \fermi-LAT, \swift- XRT and UVOT, OAGH, and Mets\"{a}hovi observations. LAT weekly fluxes reported in the upper panel for this period are all $TS>10$ detections. UV-to-near-IR fluxes are corrected for absorption.} \label{fig:multipanelMWlightcurves}
\end{figure*}

The weekly light curve in the 0.1-100 GeV band
for the 18-month period is presented in Figure
\ref{fig:18month_LAT_tuorla} (upper panel).
In order to
quantify the variability, the variability index as defined in
\citet{1FGLcatalog}, was computed from
%
\begin{eqnarray}
V = \sum_{i} w_{i}(F_{i} -
F_{wt})^{2}\ ,
\end{eqnarray}
where
\begin{eqnarray}
w_{i} = \frac{1}{\sigma^{2}+(f^{rel} F_{i})^{2}}\ ,
\end{eqnarray}
\begin{eqnarray}
F_{wt} = \frac{\sum w_{i}F_{i}}{\sum w_{i}}\ ,
\end{eqnarray}
%
and the index $i$ runs over all data points except the upper limits
(61 bins).  Here $F_{i}$ and $\sigma_{i}$ are flux values and the
relative statistical errors, respectively, and $f_{rel}$ is equal to
3\% of the flux for each interval, as suggested by
\citet{1FGLcatalog}.  For the 18 months, $F_{wt}$ was found to be
$17.48 \times 10^{-8}$ \latflux and V = 193 with 60 d.o.f., which is
consistent with variability (the probability the source is
non-variable is less than $10^{-15}$).  The same values computed just
for the first 9 weeks, containing the period of the MW campaign, give
$F_{wt} = 10.67 \times 10^{-8} $ \latflux and $V = 0.6$ with 8 d.o.f.,
which is consistent with a non-variability hypothesis at $>99$\%
confidence. Each of the LAT flux data points in the upper panel of Figure \ref{fig:multipanelMWlightcurves} has $TS>10$.

This interval
with no observed $\gamma$-ray variability characterizes the intensive
coordinated campaign period.  After the campaign, starting from about halfway
through 2008 November until 2010 February 4, BL~Lac showed variable
weekly $\gamma$-ray flux with $V =162$ with 51 d.o.f., $F_{wt }= 19.15
\times 10^{-8}$ \latflux.  In particular, a 1-week $\gamma$-ray flare
occurred in the week MJD 54928.5-54935.5 (2009 Apr.\ 7-14), with a
flux value of $72\times 10^{-8}$ \latflux.  This is about 4 times
greater than the mean flux value ($F_{wt}$) computed above. Moreover
this is the highest weekly flux yet detected from BL~Lac, although
another likely flare was detected in the week around MJD 55221
\citep[2010 Jan. 25,][]{sokolovsky10a}.  After the 2009 April flare
the source also showed slightly higher activity and flux.

The 78 points of the 18-month light curve were also used to compute the global
normalized excess variance, as defined in
\citet{vaughan03,lat-lbasvariability}: $\sigma^{2}_{NXS} = (S^{2} -
\bar{\sigma^{2}_{err}})/\bar{x}^{2}$, where $S^2$ is the variance of
the light curve and $\sigma_{err}^2 =\sigma_{i}^2+ \sigma_{sys}^2$,
the sum in quadrature of the statistical uncertainty of the flux in
the time bin and the systematic error estimate ($0.03 {\langle F_i
\rangle}$). This quantity measured an intrinsic variability amplitude
of $0.23 \pm 0.03$, in good agreement with the value found in \citet{lat-lbasvariability} reporting the weekly light curve of BL Lac (flux $E>300$ MeV) during the first 11 months.
This confirms again that BL Lac was variable after the first couple of months of \fermi science operations.
%
%
\section{Observations and results from the multifrequency campaign}\label{sect:MultifrequencyCampaign}

The 48-day PIC on BL~Lac involved the participation of the {\em RXTE}
and {\em Swift} X-ray satellites, and of ground-based radio and
optical observatories.  These included the Mets\"ahovi 13.7~m
radio-telescope operating at 37 GHz in Finland; the Very Long Baseline
Array (VLBA) in the USA, which took a multi-waveband flux-structure
observation on 2008 September 2; two telescopes of the Tuorla Observatory, Finland,
and two telescopes of the Steward Observatory, USA, for
the long-term and single-band optical monitoring as presented in Figure \ref{fig:18month_LAT_tuorla};
and the 2.1~m optical telescope of the Observatorio
Astrof\'{\i}sico Guillermo Haro (OAGH, Mexico) operating in
near-infrared, and the 1.3~m McGraw-Hill optical Telescope of the
MDM Observatory (Arizona, USA), for further multi-band optical
snapshots during the campaign period (Table \ref{table:tableobservations}).
%
\begin{table}[t!!]
\scriptsize
%
\hskip -2cm
\caption{BL~Lac coordinated multifrequency campaign (PIC) and long-term monitoring observations}\label{table:tableobservations}
%
\hskip -0.5cm
\begin{tabular}{lccc}
\hline \hline %
PIC (48 days) &  \\ %
Instrument & Energy Range & 2008 Epoch range & \# obs. \\ %
\hline %
VLBA & 4.6 - 43.2 GHz & Sep 2 & 7 \\
Mets\"ahovi & 37 GHz &  Aug 20 - Oct 6 & 22 \\
OAGH & J H K & Sep 6 - Oct 6 & 18 \\
MDM  & U B V R I & Oct 6 - 10 & 15 \\
\swift-UVOT &   W2 M2  W1 U B V & Aug 20 - Oct 2 & 141 \\
\swift-XRT &  0.4-8  KeV & Aug 20 - Sep 18 &  24 \\
\rxte-PCA  & 3-18 KeV  &  Aug 20 - Sep 8  & 19 \\
Fermi-LAT  &  100 MeV - 100 GeV &  Aug 19 - Oct 7 &  48 days \\
\hline \hline %
18 months &  \\ %
Instrument & Energy Range & Epoch range (MJD) & \# obs. \\ %
\hline %
Tuorla  & R & 54709.8 - 55191.8 & 162 \\
Steward & V & 54743.2 - 55213.1  & 89 \\
Fermi-LAT  &  100 MeV - 100 GeV &  54682.7 - 55070 &  78 weeks \\
\hline \hline %
\normalsize
\end{tabular}
\end{table}

\normalsize
%
%
\subsection{X-ray and UV observations and results}\label{subsect:swift}
%
The {\em RXTE}/PCA space observatory performed sub-daily monitoring
of BL~Lac for 20 days, during the same epoch as LAT and \emph{Swift}, for a
total of 80 pointings between 2008 August 20 and September 9.  Only
PCU2 was active, for a total exposure of 157.3 ks.  The PCA STANDARD2
data were reduced and analyzed with the routines in HEASOFT V6.8 using
the filtering criteria recommended by the \rxte\ Guest Observer
Facility.  Only the top-layer events were processed, and a check was
performed in the 40-100 keV range to assure that the model background
reproduced the observed one.  The average net count rate in the 3-18
keV band is $0.55\pm0.01$ cts\,s$^{-1}$\,pcu$^{-1}$.  The \rxte\
spectra were extracted and fitted separately for each pointing, and
summed to obtain the average spectrum.  Each spectrum is
well-fit by a single power-law model with Galactic absorption.  The
Galactic $N_{\rm H}$ was fixed at two values: that measured by
\citet{elvis89}, based on dedicated 21-cm observations
($2.015\times10^{21}$ cm$^{-2}$); and the sum of this value with that
inferred from millimeter observations \citep{lucas93}, which includes
the contribution from molecular hydrogen along the line of sight
($N_{\rm H}\simeq 3.6\times10^{21}$ cm$^{-2}$).  In the \rxte\ band,
however, this difference yields negligible effects, with a
difference in spectral indices $\Delta\Gamma_{X}\sim 0.02$.

The \rxte\ data showed modest flux variations, with a rate which
remained constant at $\sim0.6$ cts\, s$^{-1}$ up to MJD 54708, and
slowly decreasing to $\sim0.4$ cts\, s$^{-1}$ towards MJD
54718. Spectral variations were modest as well, ranging from
$1.95\pm0.09$ to $2.28\pm0.07$.  The total average spectrum is
well-fit with a single power-law model, with $\Gamma_{X}=2.18\pm0.06$,
and an unabsorbed flux in the 2-10 keV band of $F_{2-10}=(7.2\pm0.3)
\times 10^{-12}$ erg\,cm$^{-2}$\,s$^{-1}$ and $\chi^2_{r}=0.81$ (15
d.o.f.).  This spectrum can also be fit by a concave log parabola
or broken power-law model, with the slope hardening above 8-9 keV from
$\sim2.2$ to $\sim 1.9$ ($\chi^2_{r}\simeq 0.7$); however, the low
statistics do not allow one to determine if the improvement is significant, and
the single, rather flat  ($\Gamma_{X} \simeq 2$, therefore $\nu F_{\nu}
\propto \nu^0$), power law gives a good fit.

In the context of the overall SED, the low X-ray flux and the
$\Gamma_{X} \simeq 2$ spectrum (flat X-ray SED) is evidence that the \rxte\ band might correspond
to the passage between the two humps of the blazars' SED, as typical
for ISPs. In the following we show that the \swift-XRT
results provide us further evidence for this conclusion.

%
\begin{figure}[t!!]
\centering
\vskip 0.3cm
\hskip -0.6cm
\resizebox{9cm}{!}{\rotatebox[]{-90}{\includegraphics{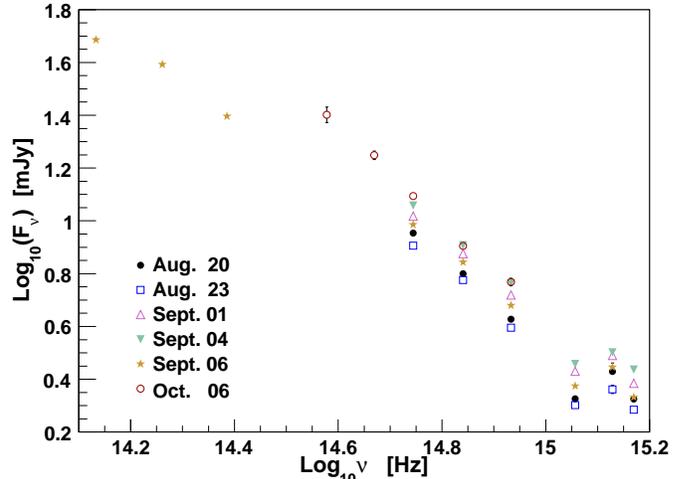}}}
\vskip -1.5cm \caption{Simultaneous near-IR to UV SEDs of BL~Lac at different epochs, collected with UVOT (
Aug. 20, 23, Sept. 1, 4), OAGH (Sept. 6), and MDM (Oct. 6) observations.} \label{fig:nearIRtoUVSEDflux}
\end{figure}
%
%
\begin{figure}[t!!]
\centering
\resizebox{\hsize}{!}{\rotatebox[]{-90}{\includegraphics{Fig6a_RXTE_powspec.ps}}}\\
\resizebox{\hsize}{!}{\rotatebox[]{-90}{\includegraphics{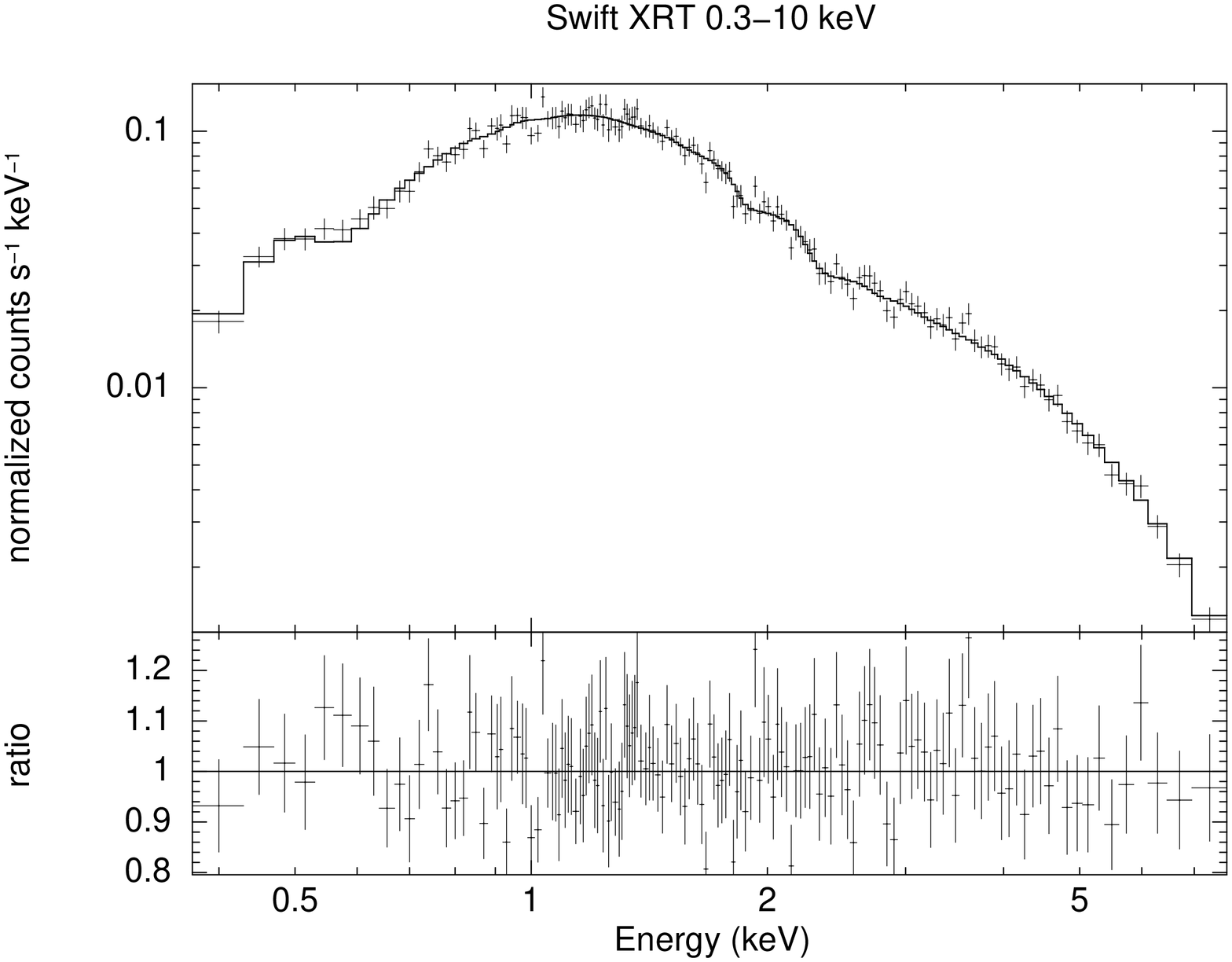}}}
\vskip -0.3cm \caption{\emph{Top panel:} \rxte\ combined 3-18 keV spectrum of BL~Lac extracted cumulating the 20 days of observations of this X-ray satellite during the campaign. \emph{Bottom panel:} \swift-XRT combined 0.4-8 keV spectra of
BL~Lac extracted cumulating the 24 days of observations of this X-ray satellite during the campaign.} \label{fig:XRTspec}
\end{figure}
%
%
The \swift\ gamma-ray burst satellite \citep{gehrels04} performed a
daily monitoring simultaneous to \rxte\ and LAT observations for 20
days (from 2008 Aug.\ 20, 15:19 UT, to 2008 Sep.\ 09, 02:37 UT;
MJD 54698.64 - 54718.11), plus 3 more separated observations after 18
September.  Data reduction and analysis was performed running a
processing script customized for the XRT and UVOT data (Donato et al.,
in preparation).  The script reprocesses the \swift\ data, stored in
the HEASARC archive using the standard HEASoft software (version 6.8)
and the latest calibration database (20091130 for XRT and 20100129 for
UVOT).  The reduction of XRT data consists of running xrtpipeline,
selecting only the events with 0-12 grades in photon counting mode
(PC).  The UVOT image mode data are processed following the steps
reported in the UVOT Software Guide 2.2.

The XRT data analysis is performed using the script \texttt{xrtgrblc},
available in the HEASoft software package. In brief, the script
selects the best source and background extraction regions based on the
source intensity (a circle and an annulus, respectively). In the case
of BL~Lac, the source extraction region has typically a 55\arcsec
radius and the background region has a 110\arcsec-210\arcsec
inner-outer radius. Field sources are excluded applying circular masks
whose radius depends on the field source intensity.  Using these
regions, source and background count rates, spectra and event lists are
extracted.  The net source count rate must be corrected for
irregularities in the exposure map and the Point-Spread Function
(PSF).  The total correction factor is obtained using the HEASoft tool
xrtlccorr.

The script that handles the UVOT analysis is \texttt{uvotgrblc} (available within HEASoft).
It determines the presence of field sources and excludes
them from the background region with
circular regions whose sizes depend on their intensity. The optimal
background region is chosen after comparing 3 annular regions centered
on the main source in the summed images. The region with the lowest
background is selected. The inner and outer radii are 27\arcsec and
35\arcsec, respectively, for all the filters except $V$, for which the
two radii are 35\arcsec and 42\arcsec. The source extraction region is
also intensity dependent: The script selects the aperture size between
3\arcsec and 5\arcsec based on the observed source count-rate and the
presence of close field sources.  BL~Lac is a relatively bright object
and an extraction region of 5\arcsec is preferred. The field of view
in the $V$ and $B$ filters is full of field sources and in particular
there is a bright star east of the blazar. To avoid contamination from
such a source, the extraction region is reduced to 3\arcsec.  The
script estimates the best source position using the task \texttt{uvotcentroid}
and the photometry using the task \texttt{uvotsource}. The obtained values are
already corrected for aperture effects while we used the dust maps of
\citet{schlegel98} and the Milky Way extinction curve of \citet{pei92}
to compensate the Galactic extinction. XRT and UVOT light curves and
SEDs are shown in Figures \ref{fig:multipanelMWlightcurves},
\ref{fig:nearIRtoUVSEDflux}, and \ref{fig:XRTspec}.

As with the harder energy band with \emph{RXTE}, BL~Lac does not
show any major sign of variability in the \emph{Swift}-XRT data. The same
Galactic ($N_{\rm H}\simeq 3.6\times10^{21}$ cm$^{-2}$) fixed value
was used also for the XRT analysis. The 2-10 keV flux derived from
\swift\ observations lies between 6 and 8 $\times 10^{-12}$ \flux with
a hint of a lower intensity at the end of the campaign when the flux
was below $5 \times 10^{-12}$ \flux. We found that in the day-by-day
spectral analysis the photon index does not change significantly
either. Using a single power-law model with fixed Galactic absorption,
Leiden/Argentine/Bonn (LAB) Survey of Galactic HI
\citep{kalberla05} weighted average equal to $1.73\times 10^{21}$ cm$^{-2}$), the
slope varies between 1.75 and 2.06. Since the typical error on this
parameter is of the order of 0.1, the variability is within 2 sigma.

The absence of spectral changes allows us to extract an accumulated
spectrum. As a first step, a comparison of the XRT and
\rxte\ spectra accumulated over the same period and using the same
lower energy threshold of 3 keV results in very good agreement with
the power-law photon index $\Gamma_{X}$ found by \rxte.  The best fit
in the 3-8 keV range is obtained with a power-law whose slope is
2.19$^{+0.13}_{-0.11}$. The 2-10 keV integrated flux is
$6.37^{+0.12}_{-0.11} \times 10^{-12}$ \flux. The X-ray photon index
is consistent within $1 \sigma$ with the results obtained using \rxte\
data.
%
%
\begin{figure*}[tttttbh!!]
\centering \hskip -0.4cm%
\resizebox{15cm}{!}{\rotatebox[]{-90}{\includegraphics{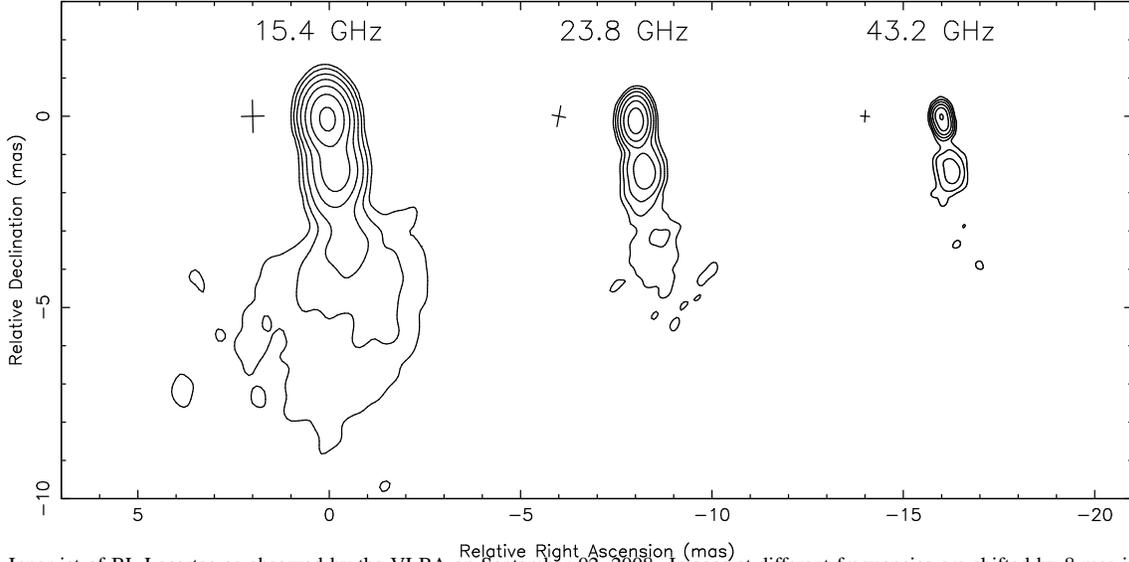}}}
\vskip -4.0cm
\caption{Inner jet of BL~Lacertae as observed by the VLBA
on September 02, 2008.  Images at different frequencies are shifted by 8
mas in relative Right Ascension.  For the 15.4, 23.8 and 43.2 GHz image map peaks
are 1.69, 1.52, 1.32 Jy/beam,  first contours are 1.70, 3.00, 5.00
mJy/beam respectively.  The contour levels are increased by a factor of
3. Beam size (natural  weighting) for each frequency is indicated by the
cross to the left of the  corresponding image.} \label{fig:Svlba_maps}
\end{figure*}

Using the entire XRT energy band from 0.3 to 10 keV (Figure
\ref{fig:XRTspec}) and profiting from the increased statistics (the
spectrum is binned with 100 counts/bin), the spectral fit is more
constrained and underlies two emission components: the best fit
($\chi^2_r=0.82$ with 146 d.o.f.) is obtained with a broken power-law
with the energy break located at 1.71$^{+0.15}_{-0.14}$ keV and fixed
absorption at the Galactic value. The soft and hard photon indices
are 2.57$^{+0.06}_{-0.05}$ and 2.09$\pm 0.05$, respectively and
therefore the break in this XRT spectrum is significant compared to
the power law. The 0.3-10 keV integrated flux is $9.67^{+0.14}_{-0.10}
\times 10^{-12}$ \flux.

The UVOT optical-UV light curves showed some variability during the
campaign period (Figure \ref{fig:multipanelMWlightcurves}), but this
is not well correlated with the X-ray or $\gamma$-ray emission (which is
not variable).  The daily simultaneous near-IR, optical, and UV SEDs,
reconstructed thanks to UVOT, and OAGH observations (reported in Figure
\ref{fig:nearIRtoUVSEDflux}), showed no significant spectral
variability and a trend consistent with a single power law except for
the 2 higher frequency UV bands. This may be in agreement with the
hypothesis of UV excess due to thermal emission from an accretion disk
\citep{raiteri09,raiteri10}.
%
%

\subsection{Radio band flux-structure observations and results}\label{subsect:radio}

As part of an ongoing blazar monitoring program, the 37 GHz
observations were made with the 13.7~m diameter Mets\"{a}hovi radio
telescope, which is a radome enclosed paraboloid antenna situated in
Finland. The measurements were made with a 1 GHz-band dual beam
receiver centered at 36.8 GHz. The high electron mobility
pseudomorphic transistor front end operates at room temperature.  The
observations are ON--ON, alternating the source and the sky in each
feed horn. A typical integration time to obtain one flux density data
point is 1200--1400 s. The detection limit of the telescope at 37 GHz
is on the order of 0.2 Jy under optimal conditions. Data points with a
signal-to-noise ratio $< 4$ are handled as non-detections. The flux
density scale is set by observations of DR 21. Sources 3C 84 and 3C
274 are used as secondary calibrators. A detailed description on the
data reduction and analysis is given in
\citet{terasranta98}.  The error estimate in the flux density includes
the contribution from the measurement rms and the uncertainty of the
absolute calibration.

\begin{figure}[tbh!!]
\centering
\resizebox{\hsize}{!}{\rotatebox[]{0}{\includegraphics{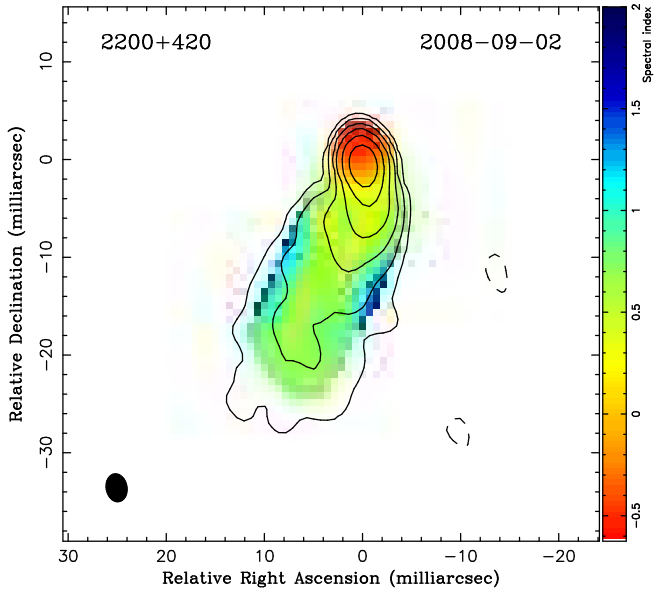}}}
\vskip -0.1cm \caption{Spectral index map ($F_\nu \sim \nu^{-\alpha}$,
$\alpha$ is shown in color) of BL~Lacertae constructed using VLBA
observations at 4.6, 5.0, 8.1 and 8.4 GHz.  The overlaid contours
represent total intensity at 8.4 GHz (the peak intensity is 1.75
Jy/beam, the first contour is 0.70 mJy/beam, the beam size is
1.57$\times$1.22 mas at PA 10.4$^\circ$.). The spectral index map was
smoothed by a median filter with radius equal to the indicated beam
size. The map shows an optically thin jet with $\alpha \sim 0.7$ and
the self-absorbed core region ($\alpha < 0.0$). A 2D cross-correlation
technique using the optically thin part of the jet was employed to
align images at different frequencies allowing reliable extraction of
the spectral information. The spectral steepening towards the jet
edges visible on the spectral index map occurs on the angular scale
comparable to the beam size and is likely an effect of the uneven
$uv$-coverage at different frequencies.} \label{fig:spctral_index_map}
\end{figure}

BL~Lac was also observed with the Very Long Baseline Array (VLBA) at
seven frequencies (4.6, 5.0, 8.1, 8.4, 15.4, 23.8 and 43.2 GHz) in the
framework of a survey of parsec-scale radio spectra of twenty
$\gamma$-ray bright blazars \citep{sokolovsky10b}. The multifrequency
VLBA observation was carried out on 2008 September 2. The data
reduction was conducted in the standard manner using the AIPS package
\citep{greisen90}. The final amplitude calibration accuracy is
estimated to be $\sim 5$\,\% at 4.6-15.4 GHz range and $\sim 10$ \% at
23.8 and 43.2 GHz. The Difmap software \citep{shepherd94} was used for
imaging and modeling of the visibility data.

The VLBA images (Fig.~\ref{fig:Svlba_maps}, and \ref{fig:spctral_index_map}) reveal a wide,
rather smooth, curved jet extending $\sim 50$ mas (at 5 GHz)
south-east from the bright compact core. A few distinct, bright
emission features aligned along the south-southwest direction can be
seen in the inner jet at higher frequencies. We have modelled the
observed brightness distribution with a small number of model
components having two-dimensional Gaussian profiles. The core region
(i.e., the bright feature at the north end of the jet) is elongated in
the 43\,GHz image and can be modelled by two Gaussian components
separated by 0.25 mas (0.32 pc projected distance). At a distances of
1.5 mas and 3.4 mas from the core, there are two other distinct
emission features in the jet. A table listing the parameters of the
distinct features (``components'') can be found in
\citet{sokolovsky10b}.

The inner 0.25 mas part of the jet (i.e., the ``core'') may contain
two (or more) distinct emission features or it may be a continuous
emission region.  The angular resolution even at 43 GHz is not
sufficient to distinguish between these possibilities. However, it is
evident that the radio spectrum is changing along this region and it
cannot be described by a single, uniform, self-absorbed, synchrotron
emitting component. A turnover caused by the synchrotron
self-absorption is detected in the averaged core spectrum at a
frequency of $\sim 12$ GHz (an average over the whole area with the
negative spectral index at Fig.~\ref{fig:spctral_index_map}). Due to
the inhomogeneity, the innermost component at mm-wavelengths most likely has an even higher turnover frequency, $\gtrsim 40$\,GHz. Using
the method described in \citet{sokolovsky10b}, the magnetic field,
$B$, of the core, in the frame of the relativistic jet, can be
constrained, given the Doppler factor, $\mathcal{D} = (\Gamma [ 1 -
\beta\cos\theta])^{-1}$, where $\Gamma=(1-\beta^2)^{-1/2}$ and $\theta$
here are the bulk Lorentz factor and jet angle to the line of sight,
respectively.  Assuming a Doppler factor $\mathcal{D} = 7.3$
\citep{hovatta09}, an upper limit can be placed on the magnetic field
strength in the core: $B < 3$\,G.  We note that this upper limit corresponds to a typical
value across an extended and inhomogeneous region contributing the
bulk of the radio emission in 4.6 - 43 GHz range.  The magnetic field
strength can exceed the above limit locally or in a more compact
regions hidden from sight in the observed frequency range by
synchrotron opacity.

%
\subsection{Optical and near IR observations and results}\label{subsect:otherNearIRandOptical}

During the campaign several multi-band near-infrared and optical observations of
BL~Lac were also performed from the ground by the 2.1~m
telescope of the OAGH observatory, Sonora, Mexico, and the 1.3~m
McGraw-Hill Telescope of the MDM Observatory, Arizona, USA. In addition a long-term monitoring
in optical $R$ and $V$ bands during the first 18 months of \fermi observations (Figure \ref{fig:18month_LAT_tuorla}), was performed by the
observing monitoring programs at the Tuorla Observatory (1~m Tuorla and 0.35~m KVA telescopes), Turku, Finland, and at the Steward Observatory (2.3~m Bok and 1.5~m Kuiper telescopes), USA respectively.

Optical ($UBVRI$) data were taken with the 1.3~m McGraw-Hill
Telescope (MDM observatory).  Exposure times ranged from 40~s ($R$ band) to ~120~s ($U$
band).  The raw data were bias and flat-field corrected using
IRAF. Instrumental magnitudes of BL~Lac and comparison stars B, C, H,
and K of \citet{smith85} were extracted using DAOPHOT and subsequently
converted to physical magnitudes through differential photometry.

INAOE (Instituto Nacional de Astrof\'{\i}sica, \'Optica y
Electr\'onica) operates the Observatorio Astrof\'{\i}sico Guillermo
Haro (OAGH) located in the Mexican state of Sonora. The 2.1~m telescope
of the OAGH has a current allocation of over 50 nights per semester
dedicated to the study of $\gamma$-ray sources, through optical
photometry and spectroscopy and infrared photometry. The Cananea Near Infrared Camera, CANICA,
is equipped with a Rockwell $1024\times 1024$ pixel Hawaii infrared
detector working at 75~K and standard $J$, $H$ and $K_{\rm s}$
bands.  The plate scale in CANICA is 0.32 arcsec per
pixel. Observations are usually carried out in series of 10 dithered
frames in each filter. Data are co-added after bias and flat-field
corrections. CANICA observations of BL~Lac presented in this work were
made on seven epochs between MJD 54715 and 54747. The data are shown
in Figures \ref{fig:multipanelMWlightcurves} and
\ref{fig:nearIRtoUVSEDflux}, and do not show signs of variability.

The photometric and polarimetry monitoring program at the Tuorla Observatory, a division of the Department of Physics and Astronomy at the University of Turku, Finland, is performed through the 1.03~m telescope at Tuorla Observatory, and the 35cm telescope at the KVA observatory on La Palma, Canary islands, Spain. Most of the monitored sources are BL Lac objects listed in \citet{costamante02}, as  potential TeV $\gamma$-ray sources\footnote{\url{http://users.utu.fi/kani/1m/}}. All photometric measures are taken using the R-filter \citep[other details in ][]{takalo08}.

The optical spectropolarimetry and spectrophotometry of blazars at the University of Arizona,
is performed on a regular basis since the launch of \fermi using the Steward Observatory 2.3~m Bok telescope on Kitt Peak, Arizona, and 1.54~m Kuiper telescope on Mt. Bigelow, Arizona, USA.  Depending on weather conditions, BL Lac is observed nightly during each of the monthly monitoring campaigns, which typically last a week.  All of the measurements have been made with the SPOL spectropolarimeter \citep{schmidt92}.  The Steward Observatory blazar monitoring program associated with the Fermi mission is described by \citet{schmidt09} and the data are publicly available\footnote{\url{http://james.as.arizona.edu/~psmith/Fermi}}.   Details concerning the reduction and calibration of both the polarimetry and photometry can also be found in \citet{schmidt09} and references therein.  The data products include high signal-to-noise flux and linear polarization spectra spanning $\lambda = 4000-7550 \AA $ and differential V-band photometry.
%
\subsection{Multifrequency correlation and SED}\label{sect:mwsed}

\begin{figure}[ttt!!]
\centering
\resizebox{\hsize}{!}{\rotatebox[]{0}{\includegraphics{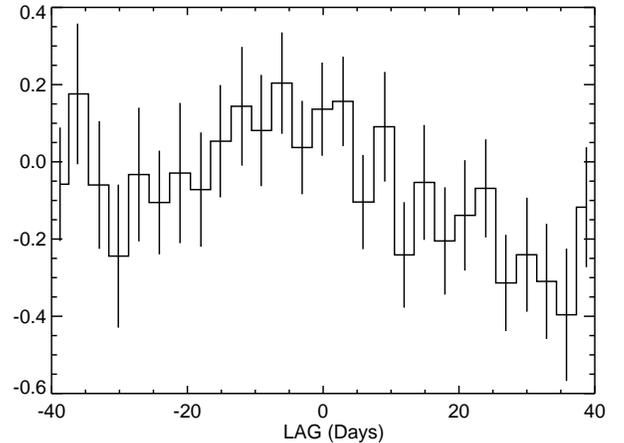}}}
\hskip -1.5cm %
\vskip -0.2cm
\caption{Discrete cross correlation function (DCCF) between the $\gamma$-ray and optical light curves of Fig. \ref{fig:18month_LAT_tuorla}.}
\label{fig:DCCF}
\end{figure}

The optical $R$- and $V$-band light curve in the bottom panel of Figure
\ref{fig:18month_LAT_tuorla} shows more rapid variability due to the
increased sampling with respect to the weekly-averaged LAT light
curve. Overlapping time intervals of the 18-month light curves
in $\gamma$-rays and in the $R$ band were used to calculate
the discrete cross correlation function (DCCF), following \citet{edelson88}.
In Figure \ref{fig:DCCF} the DCCF is plotted with error bars estimated by a Monte Carlo method,
taking measurement errors and data sampling into account \citep{peterson98}.
%
%
%
The correlation strength and lag were computed
by fitting a gaussian profile to the DCCF between
lag -20 to +20 days. The result was a correlation intensity
of $0.17 \pm 0.09$ (corrected for the effect of measurement noise)
and a lag of $-5 \pm 5$ days, where negative lag means $\gamma$-rays lagging the $R$ band.
The choice of lag range for the fit, from -15, +15 to -30, +30, as well as bin size, 3 or 4 days,
only had a small effect on the estimated time lag of the peak
and on the correlation strength, with the latter ranging from
0.14 to 0.17. The rather weak long-term correlation between
gamma-rays and optical tends to disfavor a one-zone
synchrotron plus synchrotron self-Compton (SSC) model, since in this model
one would expect electrons with approximately the same energies to
make both the optical and $\gamma$-ray emission, and might favor an
external radiation Compton (ERC) or multi-zone synchrotron/SSC model
\citep[e.g.,][]{LATPKS2155,boettcher09}.
%
\begin{figure*}[tthhhh!!]
\centering
\vskip -0.3cm %
\resizebox{16.5cm}{!}{\rotatebox[]{0}{\includegraphics{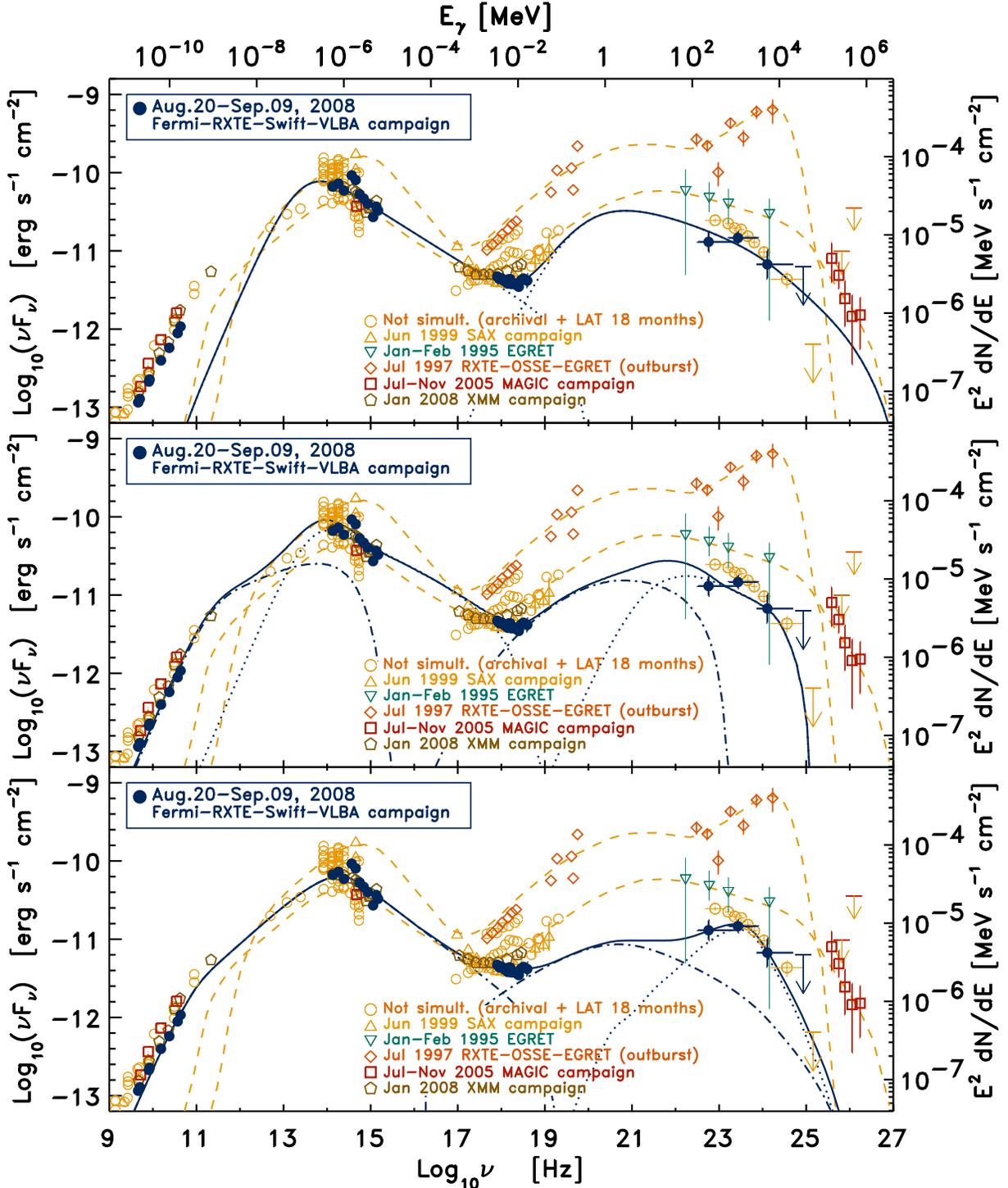}}}
\vskip -0.3cm
\caption{2008, Aug.\ 20 - Sep.\ 09 simultaneous SED of
BL~Lac assembled with data of the coordinated campaign, and some
archival data from the literature for comparison. The data are modeled
with a single zone pure-SSC (top panel), two-zone pure SSC description (center panel), and a single zone SSC plus ERC descriptions (bottom
panel).  In the two-zone pure SSC description the blue/dark dotted line
represents the compact, faster and high-energy emission component
while the blue/dark dashed-dotted line is the larger and diluted zone emission
component. In the SSC plus ERC model the blue/dark is the ERC emission (disk emission
negligible here) component and the blue/dark dashed-dotted line is the synchrotron
and SSC emission component. Past published and archival data are taken
from \citet{boettcher00,ravasio02,boettcher03,albert07,raiteri09}, and fit models for two of these past epochs \citep[the 1997 outburst campaign and the 2005 campaign, ][]{ravasio02,albert07} are represented as orange/clear long-dashed lines in each panel.}
\label{fig:SED}
\end{figure*}
%
The optical-UV variability observed during the 48 days of the
multiwavelength campaign is not correlated with the weaker X-ray
variability seen by \swift-XRT and \rxte\, nor with the non-variable $\gamma$-ray emission observed
in the same period by the LAT (Figure \ref{fig:multipanelMWlightcurves}). The
lack of correlation might support the scenario where both synchrotron
and IC photons contribute to the X-ray emission. Variability
produced by the highest energy electrons would be smeared out by the
IC component, and the mix of photons from the low and high energy
component would dilute the correlation with the synchrotron, or
synchrotron plus thermal, optical-UV emission. This would explain why variability in the optical/UV is greater than in X-rays. Moreover if the synchrotron
and IC emission components come from physically distinct regions, they
would also weaken the correlation. No evidence for a contribution by
accretion disk radiation during this low activity state of the
campaign was found, although this might contribute to the lack of
clear correlation between the X-ray and optical-UV emission.  The
recent discovery of a luminous $H\alpha$ emission line in the optical
spectra of BL~Lac \citep{vermeulen95,corbett96,corbett00,capetti10}, as likely a product of the broad
line region photoionized by the disk radiation, can suggest a more
relevant role played by the disk during major outburst with a high
$\gamma$-ray Compton dominance (as seen in the EGRET 1997 outburst).

The averaged SED from the MW campaign(2008 Aug. 20 -- Sep. 9) is
reported in Figure \ref{fig:SED}.  For the first time the
multi-frequency emission for the eponymous blazar source BL~Lac is
mapped from radio to GeV $\gamma$-rays during a low-activity state.
The relevant archival observations of BL~Lac are also reported in this
figure for comparison in addition to our data. The difference in
$\gamma$-ray integrated flux with respect to the outburst state of
1997 measured by EGRET is about 15 times lower.

The XRT spectral results are in agreement with the
\rxte\ results in the common energy range 2-10 keV, and with the usual SED properties of ISP blazars. In fact the confirmation of a flat X-ray SED ( $\nu F_{\nu} \simeq const$) of moderate flux level with both instruments in such range, makes us more confident the X-rays are from a transition region between the synchrotron and inverse Compton energy components. In addiction, as reported in Section \ref{subsect:swift}, the whole 0.3-10 keV range spectrum accumulated during the campaign by \swift-XRT showed a broken power-law best fit, i.e. a steeper soft X-ray spectral slope which
should be the tail of the synchrotron emission and a flatter slope above 3 keV, which should be the first part of the IC energy component.

The LAT $\gamma$-ray spectrum
accumulated in 18-months is reported as well for comparison (open/orange circles
representing non simultaneous data in Fig. \ref{fig:SED}).

A single-zone synchrotron self-Compton (SSC) description, usually successful for TeV blazars, has been used to model also the SED of this eponymous BL~Lac in the past
\citep{ravasio02,ravasio03}, and is thus the first step in the attempt
to reproduce its broadband emission. This model \citep{finke08} is
applied to the SED of BL~Lac (blue/dark thick line on the top panel of
Figure \ref{fig:SED}).  The energetic blob is modeled with Doppler
factor ${\mathcal D}=\Gamma=26$, comoving radius $R= 5.3 \times
10^{15}$ cm, (which is consistent with a 2 hour variability time
observed in the source on previous occasions), and tangled magnetic
field of intensity $B= 0.33 $ G.  The electron distribution has the
form of a broken power law, so that $N_e(\gamma) \propto
\gamma^{-p_1}$ for
$\gamma_{min}=600<\gamma<\gamma_{break}=1.4\times10^3$, and $N(\gamma)
\propto \gamma^{-p_2}$ for $\gamma_{break}<\gamma<\gamma_{max}=10^6$
where $\gamma=E/(mc^2)$ is the electron Lorentz factor.  The best fit
electron indices were found to be $p_{1} = 2.0$ and $p_{2} = 3.8$,
which is consistent with electrons in the fast cooling regime.  It is
interesting to note that $\gamma_{min}$ is well constrained, for a
given a blob size, by the X-ray observations, and thus it is
impossible for this model to reproduce the radio points.  This means
that, if this model is correct, the emitting region must be closer to
the black hole than the radio photosphere.  In this model the Poynting
flux power and the kinetic power in relativistic electrons are
$L_{jet}^{(B)} = 1.5 \times 10^{43}$ erg s$^{-1}$ and $L_{jet}^{(e)} =
2.2 \times 10^{44}$ erg s$^{-1}$, respectively, implying an
equipartition fraction $\epsilon_B \equiv U'_B / U'_e =
L_{jet}^{(B)}/L_{jet}^{(e)}= 0.068$. This is well below the
equipartition value, which may be argued as a reason to disfavor this
model.  In this case the Doppler factor is considerably higher than the radio-variability Doppler factor
reported in \citet{hovatta09}, although they evaluated that from the estimated timescale of each radio flare component and the known brightness temperature, referred therefore to a much larger region beyond the radio photosphere than this model suggests.

Since the one-zone SSC model has difficulty reproducing the
variability of this source and giving reasonable energetics ($\epsilon_B$ substantially below 1), as further step we
attempt a fit with a two-zone SSC model (blue/dark thick line in the
central panel of Figure \ref{fig:SED}), using two emitting blobs via SSC emission: a
compact and faster emission region and a larger, slower and diluted
region accounting also for the radio-band emission. The energetic blob responsible for most
of the emitted power and originating in a limited part of the jet is
modeled with ${\mathcal D}_1=10$, $R_{1}= 9.0 \times 10^{15}$ cm,
$B_{1}= 0.45 $ G. The radio-band and hard X-ray emission is
reproduced using a much larger emitting region characterized by
parameter values ${\mathcal D}_2=6.5$, $R_{2} = 1.6 \times 10^{17}$
cm, and $B_{2}=0.02$ G. Doppler factors of the two zones are found
to be closer to the average value (${\mathcal D}=7.3$) found in previous years by
\citet{hovatta09} than the one-zone SSC model presented above.  The
kinetic partial differential equation of this model describes the
evolution of the particle energy distribution $N(\gamma,t)$ after the
injection of shock-accelerated electrons, with a rate equal to a power
law with an exponential cutoff: $Q(\gamma) \propto
\gamma^{-p}e^{-\gamma/\gamma_{max}}$, ${\rm [cm^{-3}~s^{-1}]}$, where
$\gamma=E/(mc^2)$, between $\gamma_{min}$ and $\gamma_{max}$. More details on the numerical model
can be found in \citep{ciprini08,ciprini10}.
For the compact region (1) we obtain $\gamma_{min}^{(1)} = 3.0 \times 10^{3}$
and $\gamma_{max}^{(1)} = 7.0 \times 10^{5}$ with soft injection index
$p^{(1)} = 3.5$, and $L_{jet(1)}^{(B)} = 6.1 \times 10^{42}$ erg s$^{-1}$, $L_{jet(1)}^{(e)} =
2.2 \times 10^{42}$ erg s$^{-1}$, implying an equipartition fraction $\epsilon_B^{(1)} = 2.8 $.
Considering the larger diluted region (2) we obtain $\gamma_{min}^{(2)} = 600 $ and $\gamma_{max}^{(2)} = 3.0 \times, 10^{4}$ with injection index $p^{(2)} = 2.4$, and $L_{jet(2)}^{(B)} = 1.6 \times 10^{42}$ erg s$^{-1}$, $L_{jet(2)}^{(e)} = 7.9 \times 10^{43}$ erg s$^{-1}$, implying an equipartition fraction $\epsilon_B^{(2)} = 0.020 $. In this case for the two regions the total $\epsilon_B = 2.16$, a value slightly closer to energy equipartition than the one-zone model. The hypothesis of two synchrotron components present at different distances from the nucleus of BL Lac and accounting for the optical and the X-ray spectra respectively, was
suggested in the past in \citet{ravasio03}. In our case X-rays could be considered produced from both the emitting regions (central panel of Figure \ref{fig:SED}), and this can explain the lack of evident correlation between UVOT and XRT/RXTE light curves. On the other hand the fact that the variations in $\gamma$-rays are not seen, or not resolved by the LAT, contrary to the optical-UV variable flux, tell us that the two-zone pure SSC model does not explain the multifrequency light curve of Figure \ref{fig:multipanelMWlightcurves} observed in BL Lac during the campaign, although the energetics is closer to (and above) equipartition with respect to the previous single-zone model.

SSC emission with the relevant addition of external radiation Compton
(ERC) emission was used to explain the large Compton dominance of BL
Lac during the 1997 $\gamma$-ray outburst recorded by EGRET \citep[][]{bloom97,madejski99,boettcher00,ravasio02,boettcher04}.
An accretion disk emitting component, and emission lines
were also observed several times in this blazar \citep{vermeulen95,corbett96,corbett00,capetti10}.
The SSC plus ERC hybrid model, where X-ray
emission is produced via the SSC mechanism and the GeV $\gamma$-ray
emission is produced by ERC, was used by \citet{madejski99} for this
source. This motivates its use to see if it is able to reproduce also the 2008 low state of BL~Lac.
A model fit to this SED is shown as the blue/dark thick line
of the bottom panel of Figure \ref{fig:SED}. For this model fit an
equilibrium version of the time-dependent jet model reported in
\citet{boettcher02} was used. A power-law distribution of electrons,
$Q(\gamma) = Q_0 \gamma^{-p}$, with lower and upper cut-off
$\gamma_{min}= 1.1 \times 10^{3}$ and $\gamma_{max} = 2.0 \times
10^{5}$, respectively, and injection index $p=2.85$ is injected into a
compact region of radius $R=3 \times 10^{15}$ cm, moving
relativistically along the jet, oriented at an angle $\theta_{\rm
obs}= 3.8^{\circ}$ with respect to the line of sight, with bulk
Doppler factor ${\mathcal D}=15$ and tangled magnetic field intensity
$B= 2.5$ G. The instantaneous electron injection is self-consistently
balanced with particle escape on a time scale $t_{\rm esc} = \eta_{\rm
esc} R/c$, where $\eta_{\rm esc} = 60 $ is a free parameter, and
radiative cooling through synchrotron, SSC and ERC emission. For the
ERC emission, both the direct accretion disk radiation field and
accretion disk emission reprocessed in the BLR are taken into
account, although for our fit, the BLR-reprocessed radiation
strongly dominates the radiation which is directly from the disk.  We
assume an accretion disk luminosity of $L_{disk} = 6 \times 10^{44}$
erg~s$^{-1}$, and an emitting region distance from the black hole of
$r_0=0.1$ pc. The BLR is represented as a spherical shell of
reprocessing material with radial Thomson depth $\tau_{\rm BLR} =
0.01$, located at a distance $r_{\rm BLR} = 0.21$~pc from the
central black hole.  Although the model fits the radio data well, the assumed distance from the black hole within the BLR
implies that the blazar region is well within the radio-band photosphere.
The numerical model self-consistently evaluates the energy content in
the resulting equilibrium electron distribution, and compares this
value to the magnetic-field energy density. This model gives
$L_{jet}^{(B)} = 4.75 \times 10^{43}$ erg s$^{-1}$ and $L_{jet}^{(e)}
= 3.2 \times 10^{43}$, and ratio $\epsilon_B = 1.48$.  This is closer
to equipartition than the one- and two-zone pure SSC
models described above. Considering the equipartition ratio $\epsilon_B$
the differences between the two-zones SSC and the ERC scenario do not appear particularly significant
because of the uncertainties in the estimate of this parameter, but the
uncorrelated variability can still be explained more easily in the ERC
scenario.

The parameters used for the three SED models that are shown in Figure
\ref{fig:SED} are in agreement with the limit on the magnetic field obtained from the VLBA observations, while the Doppler factors used in the two-zone SSC are
in better agreement with the value extracted from radio-band structure
historical observations.  The magnetic field and
Doppler factor could be different for the $\gamma$-ray emitting region
than for the radio core region described in section \ref{subsect:radio},
since the $\gamma$-ray emittion region could be from a
smaller region than the radio core, closer to the black hole. A ERC+SSC model can produce more complex SEDs and less correlated variability
patterns, partly because of variability in the external radiation field.
The emission region could move through an inhomogeneous
radiation field, when it leaves the BLR. Synchrotron and SSC depend on
the magnetic field, while the ERC component does not. Therefore,
magnetic field fluctuations may also produce uncorrelated variability
patterns between optical - X-rays (SSC) on the one hand and
$\gamma$-rays (ERC) on the other.
%
%
\section{Conclusions}\label{sect:conclusion}

This is the first broad-band multifrequency campaign investigating the
low activity state of BL~Lac. This eponymous source is also peculiar,
sometimes showing a rather complex SED profile, a behavior similar to an FSRQ
\citep{vermeulen95,madejski99}, and being detected also by VHE Cherenkov telescopes like MAGIC above 200 GeV.  We observed a $\gamma$-ray flux ($E>100$ MeV)
during the 48-day multifrequency campaign (planned intensive campaign, PIC) of 2008 about 15 times lower than the level measured
by EGRET during the 1997 outburst, and lower than the 1995 EGRET
observations. This value and the weekly light curve of this campaign interval indicates a quiescent $\gamma$-ray state.  The average LAT spectrum accumulated during the first 18
months of survey showed evidence for curvature. The single photon of maximum energy detected by the LAT during these months was measured at 70 GeV, with a spectrum that could extend into the VHE range. The weekly light curve is characterized by at least two mild flares on the 1-week timescale (in the middle of 2009 April and the end of 2010 January).

During the PIC the low luminosity and non-variable $\gamma$-ray state did not correspond to the lowest
luminosity state for near-IR and optical emission. No evidence for
clear correlation between the low-energy emission component (radio to
UV bands) and the high-energy emission component (X-ray to
$\gamma$-ray bands) was found. Since one would expect correlated
variability between these components in a one-zone SSC scenario, a fit
with a two-zone SSC model was attempted. This model as well is not able
to explain the variable optical and UV emission seen by \swift corresponding to the non-variable $\gamma$-ray emission during the coordinated campaign. Therefore a hybrid SSC+ERC model was better able to reproduce the SED being closer to energy equipartition and providing a possible explanation of the uncorrelated variability.

This result tells us that the SSC+ERC model is valid for BL Lac also during low brightness states (strengthening the quasar-like, FSRQ, view of this blazar). The two-zone SSC model might still be a possibility for this blazar \citep[as suggested, for example in ][]{ravasio03}. Other blazars like S5 0716+71 appear to have evidence for two synchrotron components \citep{giommi08,chen08}. In 3C 454.3, the prototypical FSRQ, a two-zone synchrotron scenario is inferred by the variability of the dust IR emission bump resolved by Spitzer \citep{ogle10}. Therefore, as suggested in the past, BL Lac, a rather special object with a possible Seyfert-like nucleus, might be considered a miniature and real FSRQ \citep{vermeulen95,corbett96,madejski99,corbett00,capetti10}.

A multi-frequency VLBA observation performed during this campaign
demonstrated that the bright, innermost part of the radio-mm jet
(``the core'') cannot be described by a single, uniform,
self-absorbed, synchrotron emitting blob, but rather an inhomogeneous
region with spatially changing spectral properties. The synchrotron
self-absorption turnover is detected in the VLBA data which allows us
to constrain the average magnetic field in the core region to be less
than 3\,G, assuming a particular Doppler factor from previous
radio single-dish observations. SED model values are in agreement with this limit,
although this is not strictly necessary since the models assume
emission from a region closer to the black hole than the radio core.

Another interesting discovery was that during the campaign, BL~Lac
continued to show a broken, concave X-ray spectrum, as seen by the
\swift-XRT. This was not confirmed by \rxte, probably because of the differing energy range and worse sensitivity of this instrument, but the photon index $\Gamma_X \sim 2$ is in agreement with this view.
We note that the \rxte\ spectrum was mostly consistent with the XRT
one.  Both instruments observed an X-ray spectrum with small
day-to-day X-ray photon index variations during this low-activity
state. The concave spectrum and lack of spectral index variation is
believed to be a typical signature of intermediate energy peaked
blazars, and we find it exists even during non-flaring states.

\section{Acknowledgments}
%
%
%

\footnotesize

The \fermi-LAT Collaboration acknowledges generous ongoing support
from a number of agencies and institutes that have supported both the
development and the operation of the LAT as well as scientific data analysis.
These include the National Aeronautics and Space Administration and the
Department of Energy in the United States, the Commissariat \`a l'Energie Atomique
and the Centre National de la Recherche Scientifique / Institut National de Physique
Nucl\'eaire et de Physique des Particules in France, the Agenzia Spaziale Italiana
and the Istituto Nazionale di Fisica Nucleare in Italy, the Ministry of Education,
Culture, Sports, Science and Technology (MEXT), High Energy Accelerator Research
Organization (KEK) and Japan Aerospace Exploration Agency (JAXA) in Japan, and
the K.~A.~Wallenberg Foundation, the Swedish Research Council and the
Swedish National Space Board in Sweden.
\par Additional support for science analysis during the operations phase is gratefully
acknowledged from the Istituto Nazionale di Astrofisica in Italy and the Centre National d'\'Etudes Spatiales in France.
\par
S.C. acknowledges funding by grant ASI-INAF n.I/047/8/0 related to \textit{Fermi} on-orbit activities.
\par This work includes observations obtained with the NASA \textit{Swift}
gamma-ray burst Explorer. \swift\ is a MIDEX Gamma Ray Burst mission led by NASA with participation of Italy and the UK. This work includes observations obtained with NASA \textit{Rossi} XTE satellite. The ASM/\rxte\ teams at MIT and at the \rxte\ Science Operation Facility and Guest Observer Facility at NASA's GSFC are gratefully thanked.
This work includes observations obtained with the Very Long Baseline Array, USA (project code BK150). The National Radio Astronomy Observatory (NRAO VLBA) is a facility of the
National Science Foundation operated under cooperative agreement by Associated Universities, Inc. This work includes observations obtained with the
14~m Mets\"{a}hovi Radio Observatory, a separate research institute of the Helsinki University of Technology. The Mets\"{a}hovi team acknowledges the support from the Academy of Finland to our observing projects (numbers 212656, 210338, and others).
This work includes observations obtained through the Tuorla Blazar Monitoring Program, carried out with the KVA telescope on La Palma, Canary Islands and the 1~m telescope at Tuorla. Tuorla Observatory is a division of the Department of Physics and Astronomy at the University of Turku, Finland.
This  work includes observations obtained through the optical monitoring of BL Lac and other blazars using the 2.3~m Bok and 1.54~m Kuiper telescopes of Steward Observatory that is supported by NASA/Fermi Guest Investigator grants NNX08AW56G and NNX09AU10G.
This  work includes observations obtained with the
2.1~m telescope of the OAGH Observatorio Astrof\'{\i}sico Guillermo Haro, in the state of Sonora, Mexico, operated by the Instituto Nacional de Astrof\'{\i}sica, \'Optica y Electr\'onica (INAOE), Mexico.
This  work includes observations obtained with the 1.3~m McGraw-Hill Telescope
of the Michigan-Dartmouth-MIT (MDM) observatory, operated by University of Michigan, Dartmouth College, Ohio State University, Ohio University, Columbia University, in Arizona, USA.
\par The Fermi LAT Collaboration extend thanks to the anonymous referee who made useful comments.

{\it Facilities:} \facility{ {\it Fermi}},  \facility{VLBA},  \facility{RXTE}, \facility{Swift}

\normalsize

\bibliographystyle{apj}
%


\end{document}